# A Non-Invasive Method for the Safe Interaction of Cities and Electric Vehicle Fleets

Bill Power
School of Electrical and Electronic Engineering
University College Dublin
Dublin
bill.power@ucdconnect.ie

Brian Mulkeen
School of Electrical and Electronic Engineering
University College Dublin
Dublin
brian.mulkeen@ucd.ie

Anthony D. Fagan
School of Electrical and Electronic Engineering
University College Dublin
Dublin
tony.fagan@ucd.ie

Robert Shorten
School of Electrical and Electronic Engineering
University College Dublin
Dublin
robert.shorten@ucd.ie

*Abstract*—Electric and hybrid vehicles are growing in popularity. While these vehicles produce less pollution, they also produce less audible noise, especially at lower speeds. This makes it harder for pedestrians and cyclists to detect an approaching vehicle. Thus, an additional system is required to detect electric and hybrid vehicles and alert pedestrians and cyclists of their whereabouts, especially while these vehicles are driving at low speeds in cities. This paper introduces one such method based on high frequency audio emissions that are present in EVs, which arise, for example, from the process of magnetostriction. Our method is tested experimentally using 4 different tests vehicles, and a preliminary EV detection algorithm is also presented.

## I. INTRODUCTION

Electric and hybrid electric vehicles (EVs and HEVs) are in a state of worldwide growth with electro-mobility increasingly recognised as the future of personal mobility [10]. The last 8 years, in particular, have seen a discernable increase in sales of EVs [1]. This change in transport has been, in part, motivated by governments around the world through financial incentives [1]. In Ireland, for example, EV and HEV ownership is encouraged through free, public charging points [1] and tax incentives[2]. This EV-growth is projected to continue into the future with Bloomberg New Energy Finance estimating that 54% of the global vehicle fleet will be electrified by 2040 [2].

EVs and HEVs are much quieter than traditional internal combustion engine (ICE) based vehicles, making their approach harder to detect, thus raising concern with advocacy groups such as the Federation of the Blind [3]. In a 2009 report released by the National Traffic Safety Administration (NHTSA) of America, it was found that pedestrians and cyclists are more likely to be involved in a collision with a HEV performing low-speed maneuvers than a traditional ICE vehicle [4]. Table I depicts the details of this report.

The sensory ability of pedestrians is typically based on visual cues and auditory information [12]. Due to the lack of engine noise in EVs, pedestrians and cyclists need additional means of being alerted when an EV or HEV is present or approaching. This paper introduces a new non-invasive solution to this growing problem. The method is based on audio emissions that are present in EVs, which arise, for example, from the process of magnetostriction. Once validated experimentally, we then use the method to develop a portable prototype device to detect EVs and HEVs.

The remainder of this is paper is organized as follows. Related works are described and reviewed in Section II and possible sources of audio emissions in EVs are described in section III. The experimental procedure and analysis used in the paper is explained in Section IV and the results of these experiments and the preliminary results for the detection system are discussed in Section V. To finish, the conclusion is presented in Section VI.

## II. RELATED WORK

Previous work with EV and HEV detection conducted at University College Dublin is described in [5]. Here, Radio Frequency IDentification (RFID) technology was used to detect cyclists and warn vehicle drivers. The RFID reader in the vehicle would scan for an RFID tag mounted on the cyclist and when the reader detected the tag it would alert the driver of the vehicle that a cyclist was near. More recently this work was expanded to switch HEVs to electric mode when a cyclist was detected. This was done to protect the cyclist from harmful vehicle emissions [13].

Table I Details of NHTSA 2009 Report [4]

| Vehicle Type | Amount | Pedestrian Collisions | % Vehicles | Cyclist Collisions | % Vehicles |
|---|---|---|---|---|---|
| HEV | 24297 | 186 | 0.8 | 116 | 0.5 |
| ICE | 1001100 | 5699 | 0.6 | 3052 | 0.3 |

---

[1] https://www.esb.ie/our-businesses/ecars/charge-point-map
[2] http://www.seai.ie/Grants

Within the specific context of EV detection, the introduction of artificial noise has been suggested to make pedestrians and cyclists aware of EVs and HEVs[3]. This possible solution is achieved by adding a speaker to the EV or HEV that outputs the sound of an ICE. This method attempts to solve the problem by making EVs and HEVs sound the same as traditional ICE vehicles. However, this method diminishes one of the added values of having a majority electrified vehicle fleet in the future, the reduction of noise pollution due to traffic noise. Of all sources of noise pollution, street traffic is the most prevalent and can be damaging to human health [11] - a reduction in noise pollution would help reduce blood pressure and the prevalence of hypertension for the commuters and inhabitants of cities [6].

Vision-based vehicle detection has also received significant attention in recent years. This detection method involves mounting sensors such as cameras or radar-based sensors to the vehicle. These sensors send data to an on-board computer that uses recognition software to detect other vehicles when they are near [17]. A review of this method for use with driver assistance and autonomous driving is give in [16].

### III. AUDIO EMISSIONS FROM EVS

Contrary to popular belief, we conjecture that EVs give rise to audio emissions. However, these emissions cannot be detected by humans as they are outside of the frequency spectrum naturally audible to humans. Importantly, we build on the realisation that these emissions can be made detectable to humans using only inexpensive hardware (a high frequency microphone), and a suitable interaction platform (smart phone/watch). There are several potential sources for this noise, some of which we now mention.

#### A. Electromagnetic Vibrations of the Electric Motor

The windings of the EV stator and the rotor create their own electromagnetic fields. These fields counteract each other while the engine generates torque[14]. The electromagnetic forces in the motor have a radial direction with angular symmetry. The sum of these forces should be zero in a defect free machine but, due to imperfections the stator is affected by the electromagnetic forces and tends to deform[7]. This deformation is due to the radial stress tensor created by the electromagnetic forces [8] and is called ovalization under radial Maxwell pressure. The ovalization excites vibrations between the motor and its connected parts generating audio signals at the frequency of the vibrations.

#### B. Switching Noise

EVs and HEVs have power-electronic components such as switching power converters and inverters. These devices switch on and off to control the level of voltage and convert between DC and AC and switch at very high frequencies of usually up to 25kHz [7]. This switching excites vibrations that produce audio

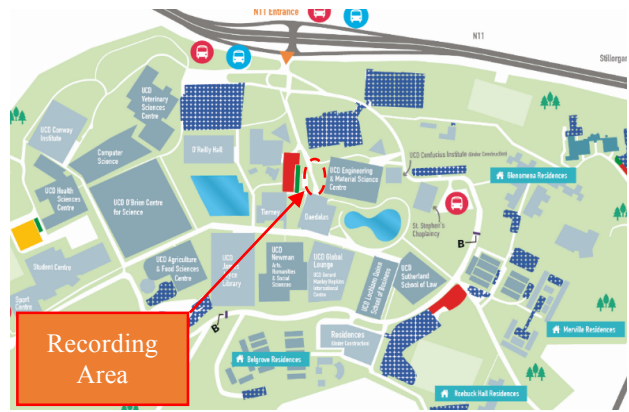

*Figure* 1. Map of University College Dublin Campus, indicating the area where each car was recorded

signals since several of these power electronic devices usually operate simultaneously.

#### C. Magnetostriction

A well-known property of ferromagnetic materials is called magnetostriction. This is the change in length of a ferromagnetic material when it is magnetized. A ferromagnetic material, exposed to an alternating field, will vibrate along its length [9]. The length of the material changes in the direction of the applied force but the volume stays approximately constant. The relative change in length can be described in equation (1) [7].

$$\lambda = \frac{\Delta l}{l}\frac{\mu m}{m} = \frac{\Delta l}{l}\lambda_s \qquad (1)$$

$\Delta l$ is the change in the length $l$, $m$ is the mass of the material, $\mu$ is the permeability of the material and $\lambda_s$ is the longitudinal change in length of the magnetically saturated material. [7]. The core of the electric motor is a ferromagnetic material, therefore, the switched magnetic field in the power converters and inverters cause a periodic deformation in the core with twice the switching frequency, since the change in length of the core is the same in both directions of the magnetic fields [7]. This causes a vibration producing an audio signal with twice the switching frequency.

### IV. EXPERIMENT SETUP

Our basic idea is that these audio emissions are detectable for a wide body of EVs. To validate this conjecture, we sampled the noise profile from a number of popular, mass market, EV and HEV models. We now briefly discuss the experimental setting in which these experiments were conducted.

#### A. Vehicles and Equipment

Table II describes the vehicles used during the experiments and Table III describes the recording equipment used. Note that mobile phone and laptop microphones are not suitable as their frequency range maximum was found to be 21kHz which corresponds to the upper limits of human audio abilities.

#### B. Main Experiment Procedure

1) Recordings were obtained for each vehicle under consistent environmental conditions. Namely, the vehicles were driven

---

[3] http://www.bbc.com/news/technology-37986774

in the same area of the university campus, at the same time of day; see Figure 1.

2) Before each experiment all microphone configurations were standardised. In addition, 30 seconds of recording data of the testing area (without the car present) was obtained

3) Each vehicle was switched to fully electric mode for the duration of each experiment.

4) *The first recording:* vehicle moving from standstill from a 1m distance from the microphone and driving in a circle with a radius of 10m, completing 2 revolutions.

5) *The second recording:* vehicle moving from standstill from a distance of 15m from the microphone. The vehicle accelerated past the microphone with its closest distance to the microphone being between 1m and 2m.

6) *The final recording:* vehicle moving from standstill from a distance of 15m from the microphone. The vehicle drives at a constant speed between 10km/hr and 20km/hr past the microphone with its closest distance to the microphone being between 1m and 2m

C. *Processing of Recordings*

Figure 2 depicts a block diagram of our data-processing system. The initial analysis was done in real-time (1), since the SeaWave recording system allows for a real-time spectrogram analysis. The recorded data was then imported into Matlab for further analysis. The real-time spectrograms were reprocessed in Matlab using the spectrogram()[4] function native to Matlab (2). An example is shown in Figure 11. The recorded signals were also divided into short time segments, where they could be assumed to be stationary. The pwelch()[5] function (again native to Matlab) was then used to generate a power spectral density (3) for the segments of interest [15].

TABLE II. Cars Used in Experiments

| Make | Model | Year | Type |
|---|---|---|---|
| Nissan | Leaf | 2016 | Fully Electric |
| BMW | i3 | 2016 | Fully Electric |
| Volkswagen | GTE | 2016 | Hybrid |
| Toyota | Prius | 2015 | Hybrid |

TABLE III. Recording Equipment Using in Experiments

| Name | Type | Details |
|---|---|---|
| Dodotronic Ultramic192k | Ultrasonic USB Microphone | • Frequency Range Maximum: 92kHz<br>• Maximum Sampling Frequency: 192kHz |
| CIBRA SeaWave 2.0 | Audio Recording Software | • Real Time Spectrogram<br>• Adjustable Sampling Frequency and FFT size |

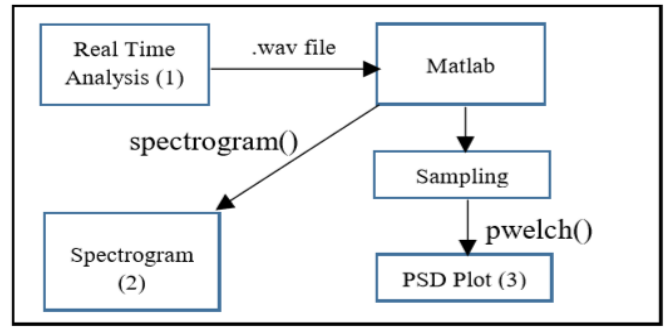

Figure 2. Recording processing block diagram

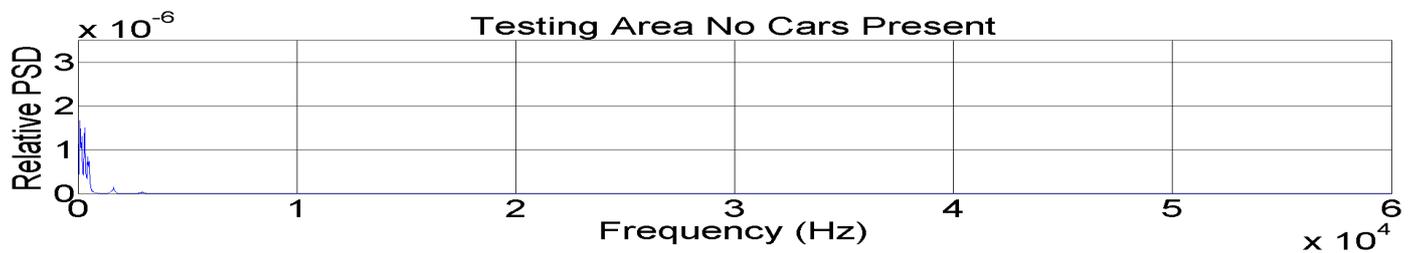

Figure 3. PSD of a the testing area with no cars present

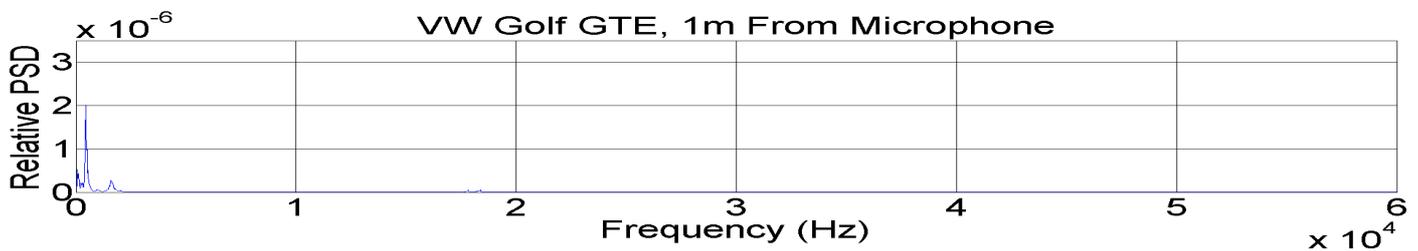

Figure 4. PSD between 0Hz and 60kHz of a VW Golf GTE driving, 1m from the microphone

---

[4] https://uk.mathworks.com/help/signal/ref/spectrogram.html
[5] https://uk.mathworks.com/help/signal/ref/pwelch.html

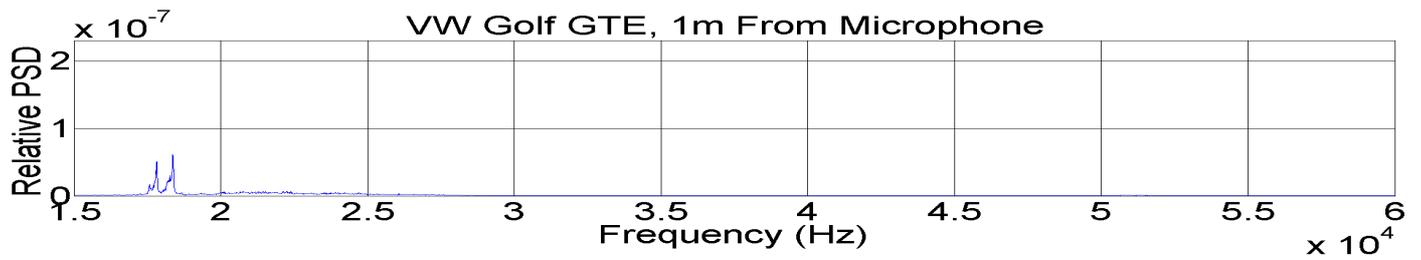
Figure 5. PSD between 15kHz and 60kHz of a VW Golf GTE driving, 1m from the microphone

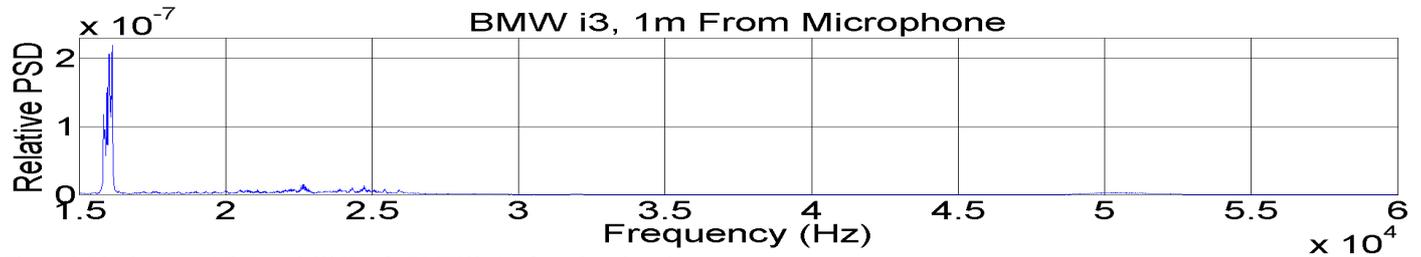
Figure 6. PSD between 15kHz and 60kHz of a BMW i3, 1m from the microphone

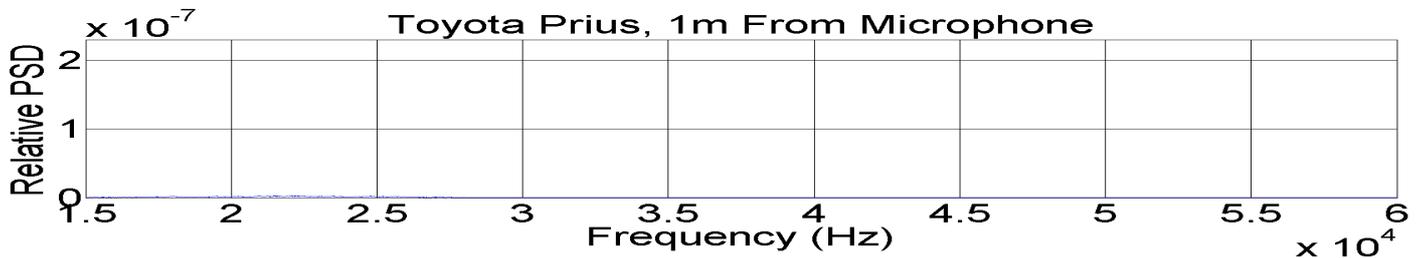
Figure 7. PSD between 15kHz and 60kHz of a Toyota Prius driving, 1m from the microphone

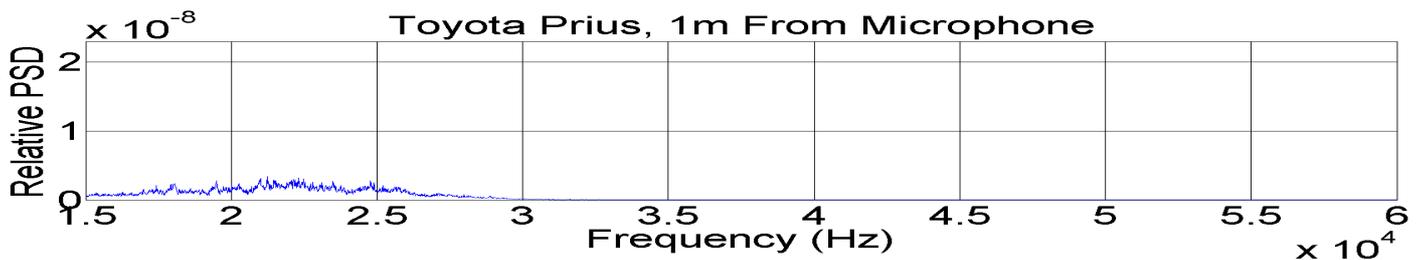
Figure 8. PSD between 15kHz and 60kHz of a Toyota Prius driving, 1m from the microphone. Y-axis limit set to $2.3 \times 10^{-8}$ Relative PSD

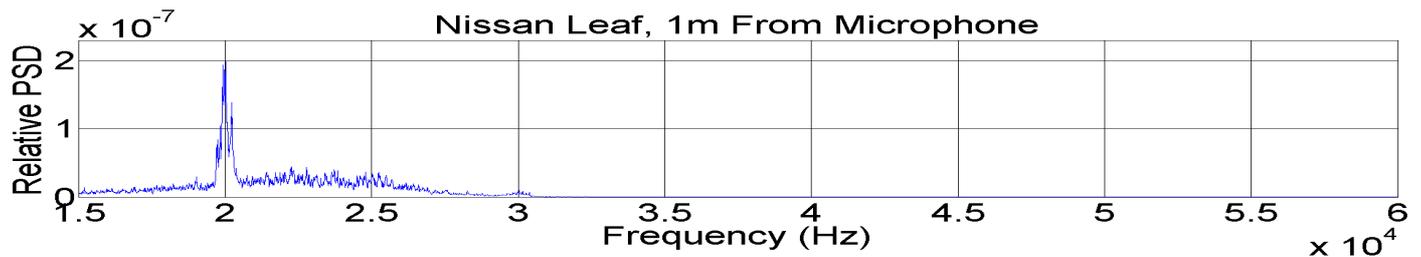
Figure 9. PSD between 15kHz and 60kHz of a Nissan Leaf driving, 1m from the microphone

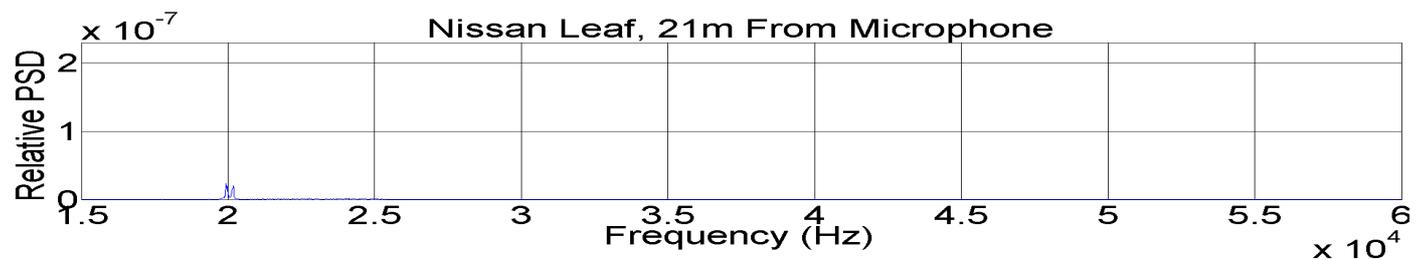
Figure 10. PSD between 15kHz and 60kHz of a Nissan Lead driving, 21m from microphone

TABLE IV. Details of Matlab Functions Used

| Name | Fs (kHz) | FFT size | Freq. Resolution (Hz) | Spectral Lines |
|---|---|---|---|---|
| spectrogram() | 192 | 2048 | 94 | 1024 |
| pwelch() | 192 | 16384 | 12 | 8192 |

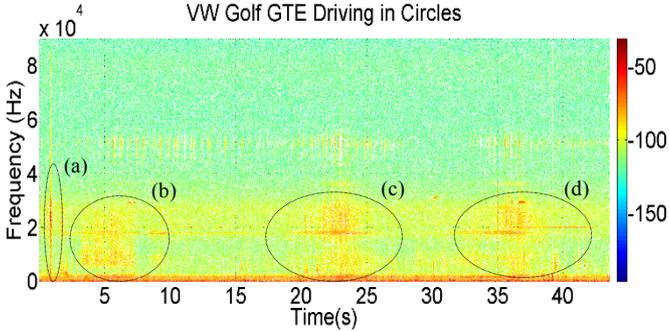

Figure 11. Spectrogram of a VW Golf GTE driving a 10m radius circles. The colour-bar on the right is in dB

## V. RESULTS

### A. Analysis of Results from Spectrogram

Figure11 shows a spectrogram of a VW Golf GTE driving in a 10m radius. The vehicle was 1m away at its closest, to the microphone, and 21m away when it was farthest away. Note in Figure 11 the low frequency background noise present in the absence of any vehicle.

(a) Here the vehicle starts. We can see a clear red line between 20kHz and 30kHz. The vehicle is then idle until 4s, where it beings to move.

(b) Now the vehicle begins to move away from the microphone. We again see some strong signals at 20kHz and strong road noise signals at the lower frequencies. The higher frequencies begin to fade as the vehicle drives further way.

(c) Now the vehicle completes its first revolution. We can see a darker yellow appear as the vehicle gets closer to the microphone. When the vehicle is within a couple of meters of the microphone we can see that the strong red line reappears at 20kHz. When the vehicle is at its closest to the microphone we can see strong signals between 20kHz and 30kHz.

(d) Between (c) and (d) we can see that the high frequency signals fade and are at their faintest at the half way point between (c) and (d) when the vehicle is at its farthest distance from the microphone. We can see the same characteristics in (d) as we did in (c) as the vehicle now completes its second revolution.

From this analysis of the spectrogram in Figure 11, and spectrograms for the other vehicles, we were able to confirm that the vehicles are in fact producing high frequency signals that are inaudible to the human ear, but which can be detected using relatively simple equipment.

### B. Analysis of Results from PSD

Figure 3 and 4 show the power spectral density of the test area without and with a car. These are dominated by low-frequency sounds, and with the linear scale used, there is little evidence of the high-frequency sounds of interest.

In Figure 5 to 10, the lower frequencies have been removed and the vertical axis re-scaled, to show the power spectral density from 15 kHz to 60 kHz more clearly. There is one example for each test vehicle driving at 1 m from the microphone and, for comparison, an example of the Nissan Leaf at 21 m from the microphone.

The main high frequency signals are between 15 kHz and 30 kHz with some signals having frequencies up to 60 kHz. Table V provides a summary of these results. The Nissan produces the largest range of frequencies as well as the strongest signals while the Toyota produces the smallest range of frequencies and weakest signals; see Figure 8. The reasons for this are as yet unclear. However, one clear difference between the Prius and the other test vehicles is its much lower pure EV driving range.

TABLE V. Analysis of Car PSDs

| Manufacturer | Distance (m) | High Freq. Range (kHz) | Strongest Freq. Component (kHz) | Strongest Relative PSD Value |
|---|---|---|---|---|
| Volkswagen | 1 | 16 – 30, 49 - 54 | 18 | $6 \times 10^{-8}$ |
| BMW | 1 | 16 – 30, 48 - 54 | 16 | $2 \times 10^{-7}$ |
| Nissan | 1 | 14 - 35 | 20 | $2 \times 10^{-7}$ |
| Toyota | 1 | 15 - 30 | 21 | $3 \times 10^{-9}$ |
| Nissan | 21 | 15 - 31 | 20 | $2 \times 10^{-8}$ |

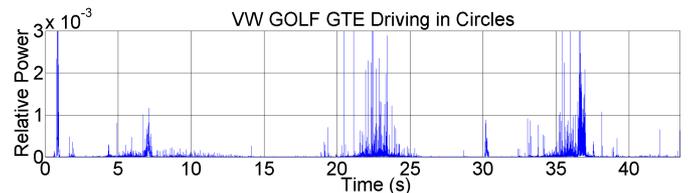

Figure 12. Relative power of a signals between 16kHz and 60kHz of a VW Golf GTE driving in circles

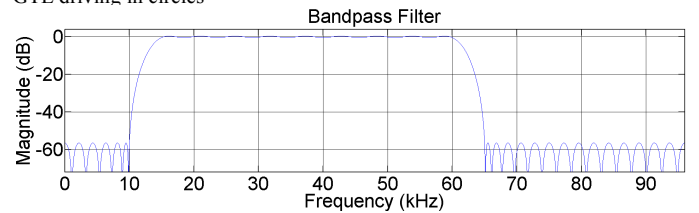

Figure 13. Bandpass filter to filter out low frequency signals and high frequency signals not produced by the car

## C. Detecion of EVs

A prototype method of vehicle detection was developed based on the previous discussion. The method of detection of an EV was as follows:

1) Collect the acoustic signal and convert it to an electrical signal.

2) Filter the signal to extract frequencies between 16kHz and 60kHz. The filter used is shown in Figure 13.

3) In each 0.5ms time segment of the filtered signal, consisting of N samples of value $x(n)$, calculate the average power using:

$$P_x = \frac{1}{N}\sum_{n=1}^{N} x^2(n) \qquad (2)$$

4) When this power exceeds a threshold power, there is a possibility that an EV is present.

5) An EV is deemed to be detected if the power exceeds the threshold in 5 out of 1000 consecutive time segments.

6) If there are 2 separate EV detection with in a 2.5s period they are combined as one detection.

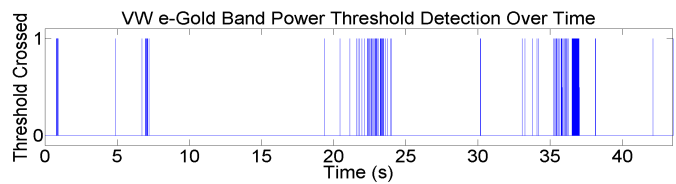

Figure14. Plot showing when the band power of the signals produced by a VW Golf GTE crosses the threshold value

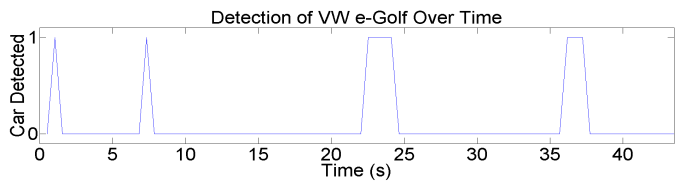

Figure15. Detection algorithm alerting when it detects a VW Golf GTE as it drives in circles

Figure 12 shows power (in the frequency band of interest) varying with time as one of the test vehicles drives in circles. Based on this, and similar plots for the other vehicles, a threshold value of $3 \times 10^{-4}$ relative power was chosen. Using this threshold, with the same data set, gives the sequence of threshold decisions shown in Figure 14. Applying the detection rule then produced the detection decisions shown in Figure 15.

Comparing Figure 15 with Figure 11 (which is based on the same data set) the algorithm can be seen detecting the car as it gets closer to the microphone, when the magnitude of the high frequency signals increases, and not detecting the car when it is farther away or when the magnitude of the high frequency signals decreases.

## VI. CONCLUSIONS

A novel EV and HEV detection method has been introduced with the aim to protect pedestrians and cyclists from unforeseen collisions with electric vehicles. The method detects high frequency acoustic signals produced by EVs and HEVs. A basic algorithm has been developed to demonstrate how these high frequency signals can be used to detect vehicles. This method has been tested using a variety of vehicles and its feasibility demonstrated experimentally. The detection algorithm needs much further refinement. This will require more tests with multiple cars in various environments to gather more data and allow the parameters of the algorithm to be optimized. Statistical analysis of large data sets is also desirable. Further, prototype detection system will be developed, suitable for use by pedestrians and cyclists.

## VII. ACKNOWLEDGEMENTS

This work was in part supported by Science Foundation Ireland grant 16/IA/4610. The authors also thank Brian Purcell, Nissan Ireland, for his support in completing this work.